# Field emission characteristics of InSb patterned nanowires

*Filippo Giubileo\*, Maurizio Passacantando, Francesca Urban, Alessandro Grillo, Laura Iemmo, Aniello Pelella, Curtis Goosney, Ray LaPierre, Antonio Di Bartolomeo\**

Dr. F. Giubileo
CNR-SPIN Salerno
via Giovanni Paolo II n. 132, Fisciano 84084, Italy
E-mail: filippo.giubileo@spin.cnr.it

Prof. M. Passacantando
Department of Physical and Chemical Science
University of L'Aquila, and CNR-SPIN L'Aquila
via Vetoio, Coppito 67100, L'Aquila, Italy

Dr. F. Urban, A. Grillo, Dr. L. Iemmo, A. Pelella, Prof. A. Di Bartolomeo
Physics Department "E. R. Caianiello", University of Salerno and CNR-SPIN Salerno
via Giovanni Paolo II n. 132, Fisciano 84084, Italy
E-mail: adibartolomeo@unisa.it

Dr. C. Goosney, Prof. R. LaPierre
Department of Engineering Physics, McMaster University
1280 Main Street West, Hamilton, Ontario L8S 4L7, Canada

InSb nanowire arrays with different geometrical parameters, diameter and pitch, are fabricated by top-down etching process on Si(100) substrates. Field emission properties of InSb nanowires are investigated by using a nano-manipulated tungsten probe-tip as anode inside the vacuum chamber of a scanning electron microscope. Stable field emission current is reported, with a maximum intensity extracted from a single nanowire of about 1 µA, corresponding to a current density as high as $10^4$ A/cm$^2$. Stability and robustness of nanowire is probed by monitoring field emission current for about three hours. By tuning the cathode-anode separation distance in the range 500 nm – 1300 nm, the field enhancement factor and the turn-on field exhibit a non-monotonic dependence, with a maximum enhancement $\beta \approx 78$ and a minimum turn-on field $E_{ON} \approx 0.033$ V/nm for a separation $d$ =900nm. The reduction of spatial separation between nanowires and the increase of diameter cause the reduction of the field emission performance, with reduced field enhancement ($\beta$<60) and increased turn-on field ($E_{ON}$≈0.050 V/nm). Finally, finite element simulation of the electric field distribution in the system demonstrates that emission is limited to an effective area near the border of the nanowire top surface, with annular shape and maximum width of 10 nm.

## 1. Introduction

Indium antimonide (InSb) is a III-V semiconductor material with a face-centered cubic structure having a small direct band-gap of 170 meV at room temperature, corresponding to the long-wave (6.2 µm) infrared (IR) regime. InSb has several applications in the field of high-speed, low-power electronics for data processing, and infrared optoelectronics.[1–3] It has also been recently demonstrated to work as a high-performance anode for lithium-ion batteries.[4]

One-dimensional InSb nanowires (NWs) have attracted growing attention in recent years due to their unique structural and physical properties towards next-generation electronic, optoelectronic and photovoltaic devices,[1,5–8] such as field-effect-transistors,[9–11] and infrared detectors.[12,13] For instance, high mobility up to 77000 $cm^2V^{-1}s^{-1}$ have been reported in InSb NW-based field-effect-transistors.[9] The use of InSb NWs to realize IR detectors working from mid- to long wavelength region has been reported.[14,15] Middle-infrared photodetectors based on a metal–semiconductor-metal structure have been fabricated using electrochemically synthesized InSb NWs, showing high stability and excellent responsivity (8.4×10$^4$ AW$^{-1}$).[15] Such enhanced properties have been explained in terms of high surface-to-volume ratio and 1D nanostructure of the photodetectors that significantly reduce the scattering and trapping phenomena, as well as the transit time between the electrodes. Multispectral optical absorptance in the short and mid-IR regions has been demonstrated[12] in top-down etched InSb NWs arrays, obtained by reactive ion etching of thin films produced by molecular beam epitaxy. In particular, it is possible to obtain highly tuneable absorptance from 1.61 to 6.86 µm.[12] Furthermore, InSb NWs with diameter below 50 nm reach the quantum capacitance limit, providing significant improvement in device performance.[16] Single crystalline *n*-type InSb NWs have also been demonstrated as gas sensors at room temperature for $NO_2$ detection down to one part-per-million.[17] Moreover, InSb is considered an environmentally friendly material with respect to other semiconductors containing As and/or P.

InSb NWs are promising candidates also for developing large area cold cathodes. It has been reported that the electron tunnelling barrier is reduced due to high carrier concentration and relevant surface accumulation layer in InSb NWs.[18] This makes InSb NWs suitable for field emission. However, few investigations have been reported about their field emission characteristics, up to now.[18–20]

The field emission is the quantum tunnelling phenomenon for which electrons near the Fermi level escape from a metal (or a semiconductor) due to the application of a strong electric field (up to $10^8$ V/cm for a flat surface) that reduces the surface potential barrier (making it triangular) at the metal(semiconductor)/vacuum interface. Consequently, one-dimensional (1D) nanostructures represent the most favourable solution to obtain strong local field enhancement due to the high aspect ratio, the prototype being carbon nanotubes (CNTs) that have been widely investigated since the 1990s as single emitters,[21,22] vertically aligned[23–25] and radially aligned[26] arrays, bundles,[27,28] and freestanding films.[29,30] Successively, in the last decade, an enormous number of 1D nanostructures (metals and semiconductors) have been characterized as field emitters, such as NWs (Si,[31] AlN,[32–34] GaAs,[35] SiC,[36,37] ZnS,[38] $Ga_2O_3$,[39,40] GaN,[41,42] and many others[43]), nanorods,[44] nanoparticles[45–47] as well as 2D layered materials (graphene,[48,49] $MoS_2$,[50–52] $WSe_2$,[53,54] and others[55–57]).

In this paper, we investigate the field emission properties of InSb NWs. The field emission experiments are locally performed by using a piezoelectric-driven metallic probe tip, with curvature radius of about 100 nm, in order to collect electrons emitted from a single NW. We directly measure the turn-on voltage, the emission current intensity and its time stability. From the I-V characteristics and fitting to the Fowler-Nordheim theory, we estimate the field enhancement factor for very small distances and the dependence on the cathode-anode separation distance. Interestingly, we observe that field enhancement factor and turn-on field have non-monotonic dependence on the cathode-anode separation distance, reporting a maximum value β=78 and a minimum turn-on field of 0.033 V/nm for 900 nm separation. By simulating the electric field distribution, we also demonstrate that the effective emitting area is limited to an annular region (10 nm wide) near the top border of the NW.

**2. Materials and methods**

**2.1. NWs Fabrication**

InSb NWs were fabricated by a top-down etching process. First, an InSb thin film was deposited by molecular beam epitaxy (MBE) on Si(100) substrate (resistivity ρ>100 Ω·cm) previously treated by HF etching (for 1 min) to remove oxide from the surface and 15 minutes of degassing at 300 °C. Next, standard electron beam lithography and reactive ion etching techniques were employed for the patterning of the film, obtaining NWs with tapered sidewalls, top diameter 400 nm ($D_1$) and 600 nm ($D_2$), and pitch 2500 nm ($P_1$) and 1500 nm ($P_2$), respectively, as shown in **Figure 1**. Details of the complete fabrication process are reported elsewhere.[12]

**2.2. Field Emission Measurement**

The field emission (FE) measurement setup is arranged inside a scanning electron microscope (SEM) chamber by using electronically nano-manipulated probes with step resolution of 5 nm. Typically, one metallic tip (tungsten) is used to contact the sample (cathode) while a second tip (anode) is closely positioned in front of the emitter to collect the electrons. Cathode-anode separation distance can be precisely estimated by SEM imaging, due to the rotating stage on which the whole setup (sample holder and probes) is mounted. A semiconductor parameter analyzer (Keithley 4200 SCS) working as source-meter unit has been connected via vacuum feedthroughs to the electrodes (probes) in order to perform FE characterization by applying bias up to ±120 V and measuring current with resolution better than 0.1 pA. Thus, FE measurements have been performed in the vacuum chamber of the SEM at a base pressure of about $10^{-6}$ mbar.

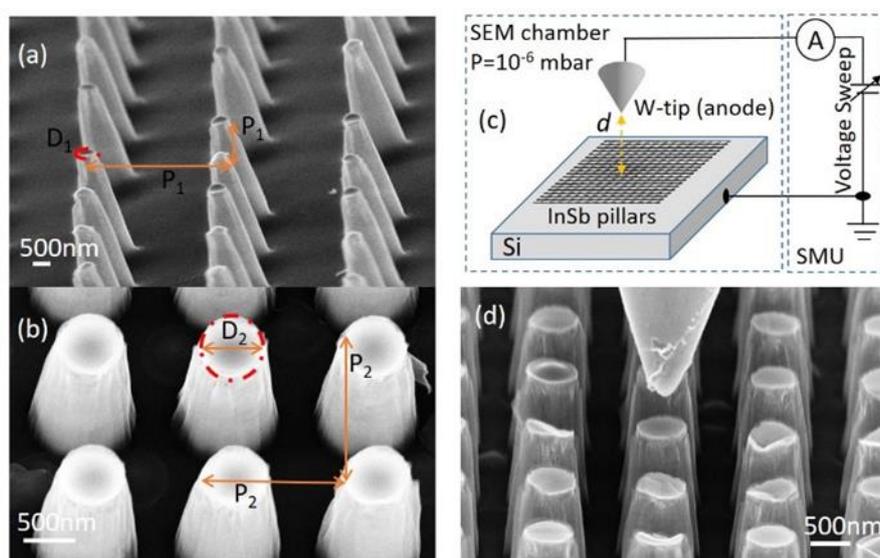

**Figure 1.** *SEM image of InSb NW arrays with (a) diameter $D_1 = 400$ nm and pitch $P_1 = 2500$ nm, and (b) $D_2 = 600$ nm and $P_2 = 1500$ nm; (c) Schematic of the field emission setup installed in the vacuum chamber of a SEM; (d) SEM image of a tungsten tip approaching one InSb NW.*

## 2.3. XPS Characterization

The surface chemical properties of InSb NWs were characterized by X-ray photoelectron spectroscopy (XPS) by means of a PHI 1257 system, working at a pressure of $10^{-9}$ mbar, equipped with a non-monochromatic X-ray source (Mg $K_\alpha$, hν=1253.6 eV) and a hemispherical analyser having a constant pass energy of 23.50 eV, corresponding to an overall experimental resolution of 0.75 eV. The analysed area was 150×150 µm$^2$ in order to investigate the chemical states of the InSb NWs array only. The survey spectrum (**Figure 2**a) clearly displays the presence of In, Sb, O, C and Si elements and, within the sensitivity of the instrument, no contamination species were observed. The binding energy (BE) for XPS spectra was referred to the C 1s core level peak (284.8 eV) of adventitious carbon[58] on the surface of the sample, assuming an energy independent BE shift. All the core level peaks, of the different elements present on the XPS regions, were deconvoluted by fitting procedure, using Voigt multipeaks and a Shirley background,[59] to obtain information on the chemical states of the NWs array. Deconvolution of the C 1s region (Figure 2b) shows a major peak at 284.8 eV related to the carbon atoms in the C-C $sp^2$ bonds,[60] used for the BE calibration. Minor peaks can be accounted for by other oxygen-containing functional groups such as C-OH centered at 285.9 eV, epoxide C-O-C at 286.9 eV, carbonyl C=O at 288.1 eV and carboxilyc C=O(OH) at 289.3 eV.[60,61]

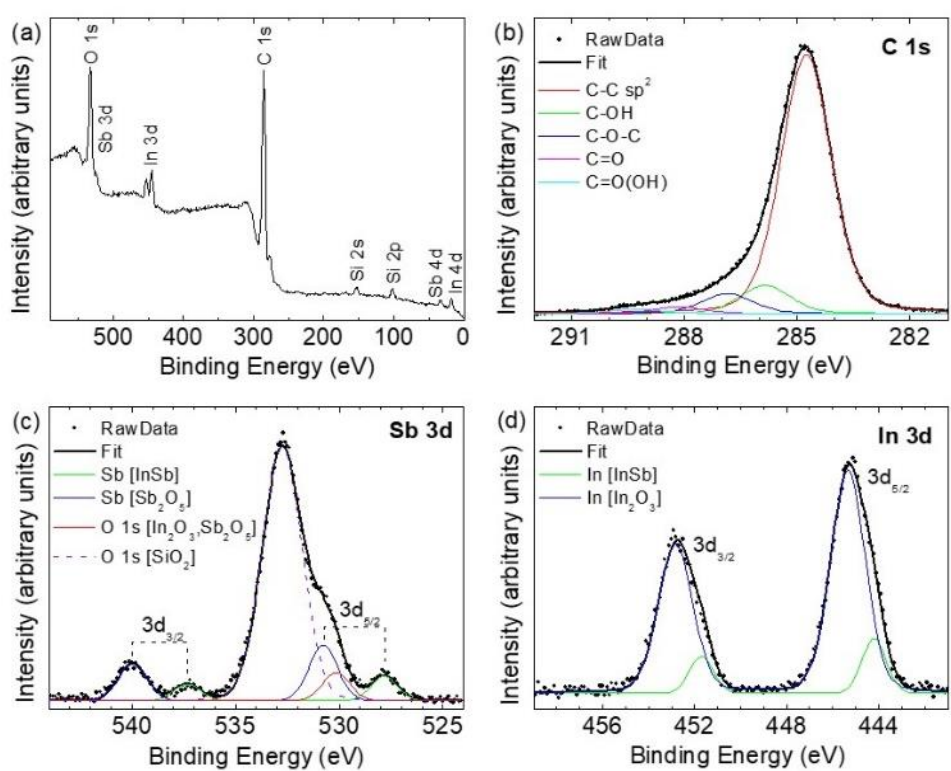

**Figure 2.** *XPS characterization of InSb NW array. (a) Survey spectrum; (b) C 1s region where C–C, C–OH, C–O–C, C=O, C=O(OH) are evident; (c) Sb 3d and (d) In 3d spectra of the InSb NW array measured with 45° take-off angle; (e) Sb 3d and (f) In 3d spectra of the InSb NW array measured with 85° take-off angle.*

## 3. Results and discussion

The Sb 3d and In 3d XPS spectra core level peaks contain two peak envelopes, $3d_{5/2}$ and $3d_{3/2}$, due to spin-orbit splitting (s.o.s.) as shown in Figure 2c and 2d. Within the Sb $3d_{5/2}$ peak envelop (Figure 2c), two differet states were deconvoluted and assigned to Sb of the InSb NWs at 527.8 eV and to $Sb_2O_5$ at 530.8 eV (i.e $\Delta E = + 3.0$ eV). In the Sb $3d_{3/2}$ peak, we resolved two states corresponding to the Sb of the InSb NWs (at 537.3 eV and s.o.s. of 9.5 eV) and to $Sb_2O_5$ (at 540.0 eV and s.o.s. of 9.2 eV).[58] The O 1s peak is overlapped with the Sb 3d peaks and it is present in two different states: one centered at 530.1 eV (corresponding to the BE of the oxygen in the $In_2O_3$ and $Sb_2O_5$ chemical states, for a native oxide of the NWs) and the other at 532.8 eV (corresponding to the BE of the oxygen in the silicon oxide, due to the Si substrate, and functional groups of the adsorbed carbon).[58]

The In 3d core level spectrum reported in Figure 2d, shows the In–Sb bonds identified at BE of 444.2 eV and 451.7 eV. These data are in agreement with the previously reported binding energies of In 3d doublet components at 444.2-444.3 eV and 451.8-451.9 eV for bulk and thin film InSb.[62,63] The contributions from $In_2O_3$ appear with a chemical shift of 1.0 eV, reasonably falling in the expected range shift 0.4-1.4 eV. [62,64,65]

A quantitative analysis of the integrated XPS peak area can provide the concentration of the elements and can provide the steochiometry of the different states evidenced in the XPS fit results. In **Table 1**, we report the atomic concentration, considering Physical Electronics sensitivity factor for In (4.359), Sb (4.477), O (0.733) and Si (0.368). Therefore, from the atomic concentration we notice that the relative amount of In and Sb are in good agreement for the formation of InSb NWs alloy. Moreover, concerning the native oxide on InSb NWs, we found that the top-most layer of the NWs is predominantely due to the formation of $In_2O_3$.

Indeed, from data in Table 1, it is evident that relatively to oxides the In concentration is significantly higher than Sb concentration (39.9% and 15.2%, respectively).

**Table 1.** *Atomic concentration calculated from XPS data.*

| Element | Atomic Concentration (%) | | | | Func. Groups |
|---|---|---|---|---|---|
| | InSb | $In_2O_3$ | $Sb_2O_5$ | $SiO_2$ | |
| In | 50.5 | 39.9 | | | |
| Sb | 49.5 | | 15.2 | | |
| O | | 44.9 | | 81.8 | |
| Si | | | | 18.2 | |

Besides, it has been demonstrated[62] that InSb sample in air undergoes surface oxidation, with a fast formation of $In_2O_3$ monolayer followed by a slower diffusion process that produces a mixture of $In_2O_3$ and $Sb_2O_5$ of about 3-5 nm thickness. According to XPS analysis, the oxidation process, after the formation of the first $In_2O_3$ monolayer, is mostly due to oxygen diffusion towards the bulk with the oxidation happening at the oxide bulk interface.

FE experiments on InSb NWs were performed at room temperature inside the SEM vacuum chamber at the base pressure of about $10^{-6}$ mbar, by using one metallic tip as cathode (directly contacting the sample surface) and finely positioning the anode (second metallic tip) at a controlled separation distance $d$ from the top surface of a NW. As standard procedure, we gently approached the anode on the sample surface and verified the current flowing in closed circuit through the NW, monitoring the current-voltage characteristic $I - V$ in the standard two-probe configuration. Soon after, by retracting the anode with controlled steps, we obtain the field emission configuration, in which the W-tip is suspended at a distance $d$ from the emitter surface. It has been already demonstrated that the tip-shaped configuration, i.e. the use of a nanometric metallic tip as anode, limits the effective emitting surface to small areas (down to $\sim 1$ µm$^2$).[66] As a consequence, the use of the a tip-shaped anode setup in this experiment gives access to the field emission from a single NW, differently from the standard parallel plate setup (for which typical mm$^2$ areas are probed).[55]

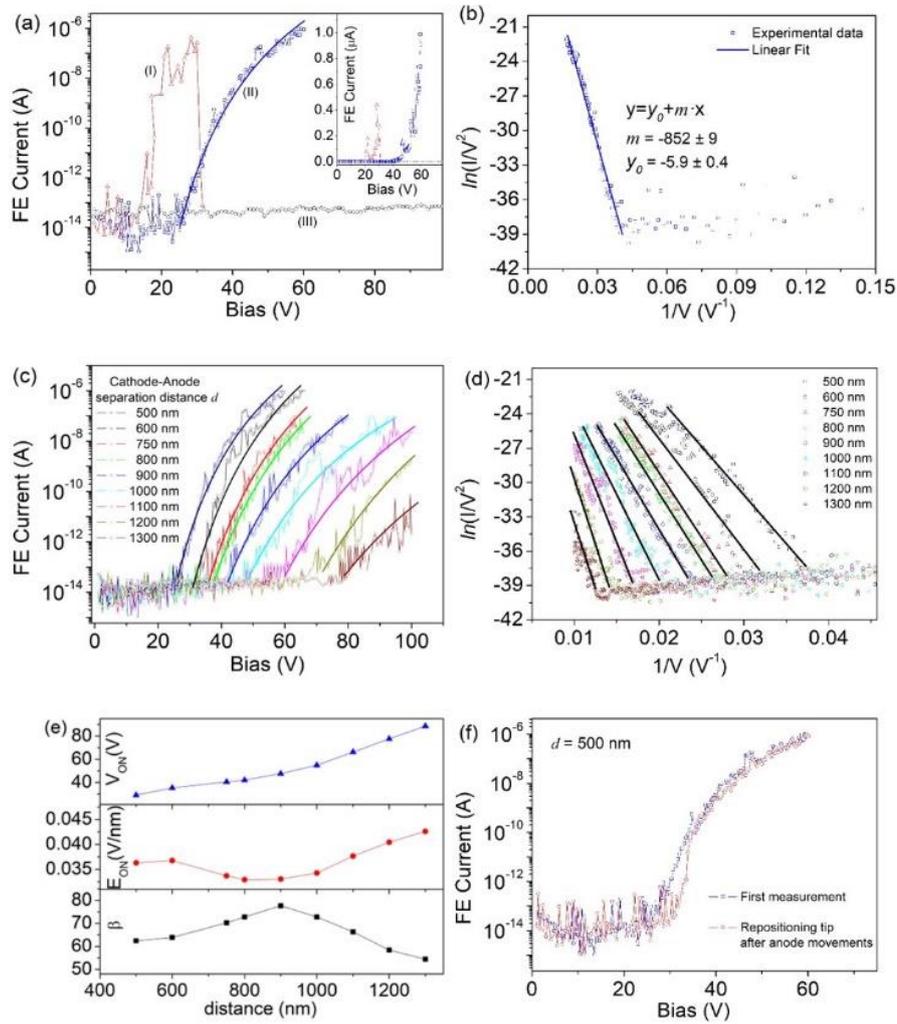

**Figure 3.** *Field emission characterization of InSb NW. (a) I − V curve measured at d = 500 nm. First sweep (I) acts as annealing/stabilization. Second (II) sweep is stable and it is compared to the situation of anode far away from the emitter (III). Inset: experimental data shown on linear scale. (b) FN-plot of curve (II) and linear fitting. (c) FE characteristics measured for several d values; data are compared to expected behaviour of FN-theory (Equation (1)). (d) FN-plots corresponding to data of Figure 3c and linear fittings (solid lines). (e) FE parameters (turn-on voltage, turn-on field and field enhancement factor) extracted from experimental data and fittings. (f) Comparison of I-V curves measured at d=500 nm before and after repositioning of the tip.*

In **Figure 3**a we show the FE characteristics ($I - V$) measured after positioning the anode tip at 500 nm separation distance above a NW. We have swept the bias from 0 to +60 V and measured the current. The first sweep (I) results in very unstable measurement with a fast increase of the current up to several hundred nA at about 15 V and large current variations up to 30 V; then for current near 1 μA there is a sudden switch-off of the FE current. This behaviour is often observed in the initial electrical sweep with irreversible modifications that result in a conditioning effect towards a stable configuration. Typically, the first sweep is enough to originate a stable emission area and rarely a further conditioning is necessary to complete the stabilization process. Mainly, the conditioning is due to the desorption of adsorbates on the surface. Such adsorbates are often responsible for nano-protrusions with reduced work function and/or increased field enhancement factor, causing FE at lower fields and with higher current density. Consequently, at large currents, local temperature can be very high, causing the evaporation of adsorbates, corresponding to the observed current drops.[67]

After the first sweep that acted as annealing/stabilization, we performed a second measurement (II in Figure 3a) in which we observe a turn-on voltage $V_{ON}$ (here defined as the voltage necessary to extract a current of 1pA) of ~25 V. The FE current exponentially raises from the setup noise level ($10^{-14}$ A) for about eight orders of magnitude up to μA. The experimental data are reported in Figure 3a using logarithmic scale to better show the sudden and fast current increase, happening in a limited bias range of 25 V-60 V. Data are also shown in linear scale in the inset of Figure 3a. We notice that due to the sub-micrometre cathode-anode separation distance, we achieve high electric fields ($10^7$-$10^8$ V/m) with a modest voltage, obtaining current densities as large as $10^3$ A/cm$^2$ (by considering as emitting surface the whole NW top surface). To analyse the FE characteristics, we will refer in the following to the Fowler-Nordheim (FN) theory.[68] The FE current $I$ can be expressed in term of externally applied bias $V$ as:

$$I = S \cdot a\phi^{-1}\beta^2 \frac{V^2}{d^2} exp\left(-b \; \phi^{3/2}\beta^{-1}\frac{d}{V}\right) \qquad (1)$$

In this formula, $S$ is the emitting surface; $a$ and $b$ are constants assuming the values $1.54 \times 10^{-6} AV^{-2}eV$ and $6.83 \times 10^9 eV^{-3/2}m^{-1}V$, respectively; $\phi$=4.57 eV is the work function of InSb;[69] β is the field enhancement factor that causes the amplification of the applied field in proximity of an apex. Thus, the local electric field $E_{local}$ can be written in term of the externally applied bias as $E_{local} = \beta V/d$. According to Equation (1), we can simulate the experimental data to confirm that the measured current is due to the FE phenomenon (simulated I-V curve is shown in Figure 3a as solid line). To extract more precise information, we follow the standard procedure to use the so-called FN-plot in which $ln(I/V^2)$ is plotted as function of $1/V$ where linear behaviour is expected from Equation (1). This is clearly demonstrated for the data of Figure 3a in the plot shown in Figure 3b. The linear fit (solid line) demonstrates the FE nature of the I-V curve. Moreover, from the slope $m$ of the linear fit we can extract β, considering that in the FN model $m = -(bd\varphi^{3/2})/\beta$. However, for a more precise estimation, we also need to take into account the tip-shape of the anode by introducing a further parameter of geometrical correction $k_{tip} \approx 1.6$.[66] Consequently, the field enhancement factor can be calculated as $\beta = (bd\varphi^{3/2}) \cdot k_{tip}/m \approx 63$. We recall here that the FN theory has been developed to describe the electron emission from a flat metallic surface and considering a potential barrier with triangular shape at zero temperature. Despite these assumptions, it has been widely demonstrated that FN theory can be correctly applied to obtain a first-approximation modelling of FE characteristics from metallic and/or semiconducting nanostructures, although corrections can be introduced, related to the finite temperature, the dimensionality of the emitters, etc. [70–74]

Taking advantage of the experimental setup inside a SEM, we have performed a complete experimental characterization of the field emission properties by varying the cathode-anode separation distance in the range 500 nm $< d <$ 1300 nm. The $I - V$ characteristics measured for different $d$ values are shown in Figure 3c and are compared to the numerical simulations according to Equation (1). We notice that the turn-on voltage (necessary to start the extraction of electrons from the InSb surface) increased monotonically by retracting the tip anode (i.e., by increasing the separation distance). For each curve shown in Figure 3c we also reported the FN-plot in Figure 3d. The linear behaviours, as evidenced by the fittings (solid lines), are a demonstration that the recorded curves originate from field emission of electrons from InSb and collected at the anode tip, positioned above the surface. The slope of the linear FN fitting can be used to extract the field enhancement factor for each value of the cathode-anode separation distance. The dependence of β on $d$ is reported in the lower plot of Figure 3e. We clearly observe a non-monotonic dependence with an increasing behaviour for small distances up to 900 nm, a maximum value β=78 obtained for $d$=900 nm and then a decreasing dependence for larger separation. Similarly, two regimes have been reported also for the FE properties of MoS$_2$ flakes,[51] showing an increasing behaviour for cathode-anode separation distance below 1μm and an inverted dependence for larger distances.

About the turn-on voltage, the upper plot in figure 3e shows the rapid increase of the voltage necessary to start the FE phenomenon for increasing distance $d$, with a minimum voltage of about 25 V at $d = 500$nm and up to 88V at $d = 1300$ nm. More interestingly, we can estimate the turn-on field $E_{ON}$ by including the correction due to the tip anode: $E_{ON} = (V_{ON}/d) \cdot (1/k_{tip})$.[66] Resulting data are reported in the central plot in Figure 3e. We observe that the turn-on field has a non-monotonic dependence on the separation distance in the range investigated. The minimum value $E_{ON} = 0.033$ V/nm is obtained for $d = 900$nm.

After the anode movement from 500 nm to 1300 nm cathode-anode separation distance, we positioned the anode back to the position $d = 500$nm in order to verify the reproducibility of our measurements. In figure 3f, we can observe that the $I - V$ curves measured at $d = 500$nm at the beginning and at the end of the experiment are very similar, confirming that the surface conditions and the field emission properties (after the annealing stabilization process) remain unaltered during the whole experiment.

The study of the field emission current over time is a further property extremely relevant to identify emitters suitable for technological applications. The emission stability for the InSb NW has been tested by positioning the anode tip at a distance of 1000 nm from the emitter and performing a measurement of the FE current at constant voltage bias of 80 V. The experimental data recorded in a time interval of about 3 hours are reported in **Figure 4**. The InSb NW is clearly a stable emitter, showing no degradation of the emitted current over the whole time interval. Moreover, the FE current has small fluctuations (standard deviation 0.03 nA) around an average emitted current of 5.60 nA.

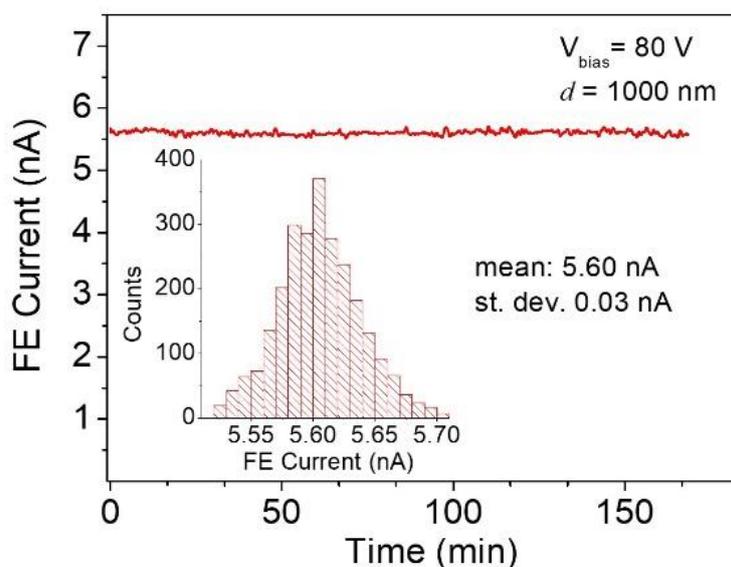

**Figure 4.** *Time stability of the FE current measured by applying a fixed bias of 80 V with the anode tip at a fixed distance of 1000 nm from the emitter surface. Inset: histogram of the measured current values.*

The FE experiment has been repeated on a different InSb NW array, having NW top surface with diameter $D_2$=600nm and pitch $P_2$= 1500 nm. The I-V curves, for three different $d$ values (800 nm, 1000 nm, 1200 nm) are reported in **Figure 5**a and compared to the FN-theory (solid lines). The linear fittings of the corresponding FN plots (Figure 5b) allowed once again extraction of the FE parameters that characterize the emitter under investigation.

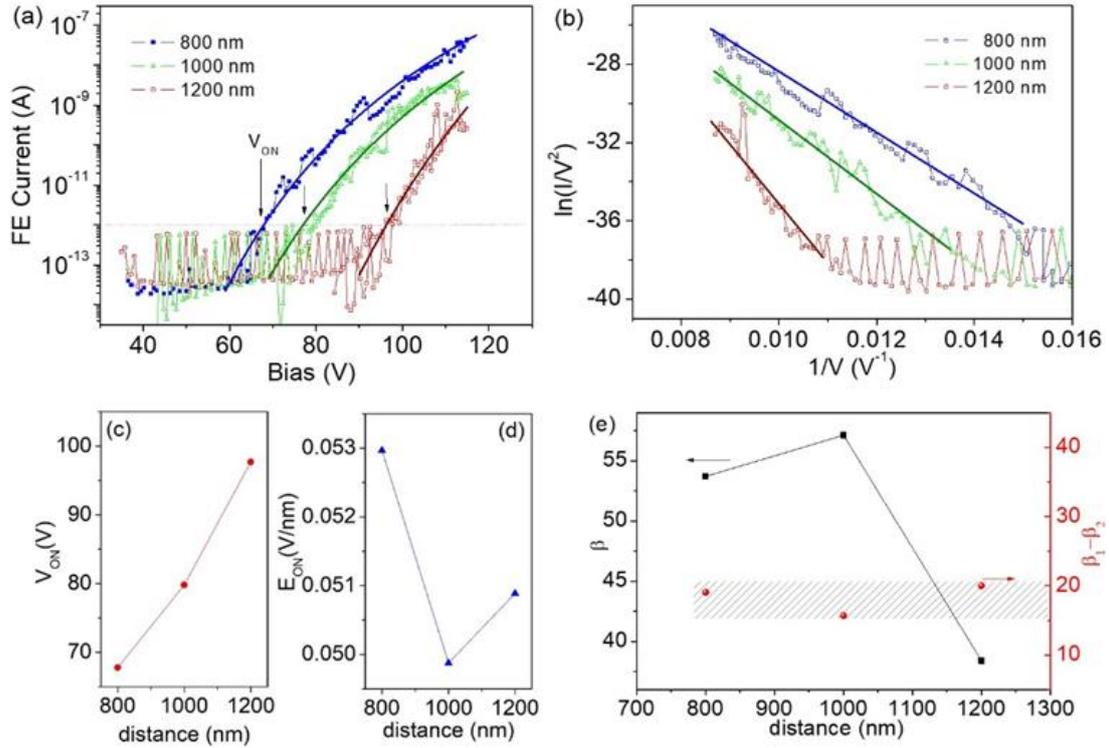

**Figure 5.** *Field emission characterization of InSb NW with diameter $D_2$ = 600 nm in the array with a pitch $P_2$ = 1500 nm. (a) I-V curves measured for three different d values of 800 nm, 1000 nm and 1200 nm; data are compared to expected behaviour (solid lines) of FN-theory (Equation (1)). (b) FN plots corresponding to data of figure 5a and linear fittings (solid lines). (c) Turn-on voltage, (d) turn-on field and (e) field enhancement factor, extracted from experimental data and fittings. Right vertical axis in figure 5e shows the difference of enhancement between the NWs in the two arrays.*

The turn-on voltage, the turn-on field and the field enhancement factor are reported in Figure 5c, 5d and 5e, respectively. We notice that the larger NW diameter ($D_2$ = 600 nm > $D_1$) and the smaller distance between the NWs ($P_2$ = 1500 nm < $P_1$) produce a higher turn-on field ($E_{ON}$ ≈ 0.050 V/nm at $d$ = 1000 nm) and a lower field enhancement factor ($β$ ≈ 56 at $d$ = 1000 nm). By comparing the β values in the range 800 nm < $d$ < 1200 nm for the two NW arrays, we observe that the enhancement $β_2$ for this second geometrical configuration is always smaller then $β_1$ (obtained for $D_1$ = 400 nm and $P_1$ = 2500 nm) with 15 < $β_1$-$β_2$ < 20 (figure 5e).

## 3.1. Electric field simulation

We performed a finite element electrostatic simulation using COMSOL MULTIPHYSICS software (Electrostatics interface – AC/DC module) to calculate the electric field for the FE system under investigation. The software solves the Laplace equation by finite element method and extracts values of the electric field at any point of the surface and/or volume of the designed model. In particular, we developed a 2D model including NWs having diameter 400 nm, height 2000 nm, and pitch 2500 nm. The tip-anode was modelled with curvature radius 100 nm, cathode-anode separation distance was varied in the range from 500 nm to 1500 nm. The voltage applied to the anode was varied between 0 V and 100 V (with a step of 10 V). Extremely fine density of the mesh was implemented in order to obtain triangular grid with a side maximum dimension of 10 nm.

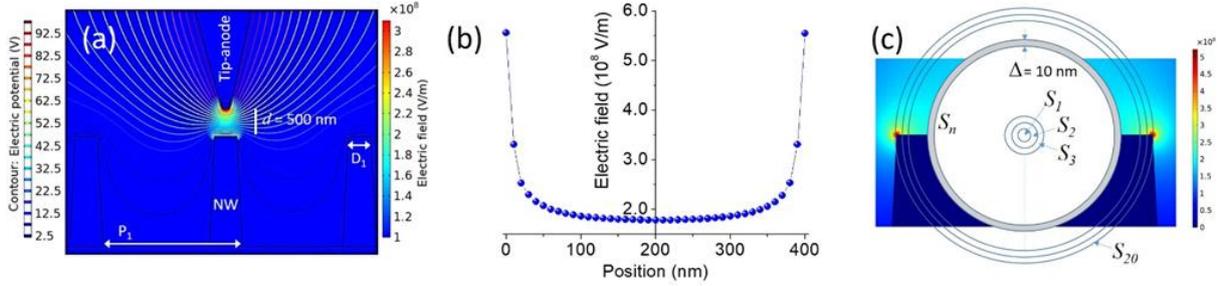

**Figure 6.** (a) M*odelling of the field emission setup, with cathode-anode separation distance d = 500 nm and bias 100 V applied on the tip-anode while grounding the cathode. Equipotential lines are shown and distribution of the electric field. (b) Electric field intensity on the top surface of the NW (along diameter). (c) Schematic of the top surface of the NW and its separation in annular areas $S_n$ with fixed width, $\Delta$=10 nm. The schematic is shown above the magnified image of the NW with the electric field distribution, showing that the enhancement of the electric field is limited to the most external annular area.*

In **Figure 6**a, we show the potential distribution (contour line) and the electric field distribution (colour surface plot) calculated for grounded cathode, 100 V bias on the tip-anode, and a cathode-anode separation distance of 500 nm. The electric field is clearly enhanced at the border of the NW top surface. The electric field profile is plotted in Figure 6b. Knowing the electric field profile, we can calculate the current according to the Fowler-Nordheim theory as:

$$I = \sum_n I_n = \sum_n S_n \cdot a\phi^{-1} E_n^2 \exp(-b\, \phi^{3/2} E_n^{-1}) \qquad (2)$$

where the current is calculated as the current emitted by any of the twenty annular areas $S_n$ which divide the emitting surface (i.e., the NW top surface) according to the step resolution (10 nm) used to compute the electric field (Figure 6c). The areas are not constant, but increase according to $S_n = \pi \cdot [(n \cdot 10)^2 - ((n-1) \cdot 10)^2]$, expressed in nm². In the computation of the total current, the calculated electric field, $E_n$, is associated with each area. We found out that almost all the FE current is emitted from the most external annular area ($S_{20}$), demonstrating that the effective emitting area is no larger than $1.2 \cdot 10^4$ nm². Consequently, for a FE current of 1 µA the current density is even higher than the previous estimation, and is of the order of $10^4$ A/cm².

From Equation (2) we also found that to obtain the real current density one should consider an effective electric field $\tilde{E}_n = \alpha E_n$, with $\alpha \approx 20$. This represents a further enhancement of the local electric field that could be explained considering that the morphology of the surface plays a crucial role in determining the field emission properties.[22,75] The emission is governed by the local field (according to Equation (2)) at the emitting sites, and consequently surface roughness and/or nano-protrusions can strongly modify/enhance the FE characteristics due to increased field enhancement around their apex. The field enhancement factor β estimated above from experimental data through the Fowler-Nordheim fittings, include both effects, i.e. the enhancement due to the NW geometry and the possible further enhancement due to existing nano-protrusions on the surface.

However, from Equation (2), we cannot exclude that the presence of In₂O₃ layer on the sample surface may significantly modify the local work-function. Indeed, In₂O₃ is a n-type semiconductor (bandgap ~2.9 eV, electron affinity ~3.7 eV)[76,77] and surface-state and/or field effect induced band bending[78] near the semiconductor surface may cause important reduction of the electron affinity on the order of (or even larger than) 1 eV.[79] For the sake of completeness, we also mention that it has been recently reported[80] that field emission from III-V semiconductors is limited by the formation of virtual electronic states localized at the surface area, causing a significant decrease of the local work function. In particular, a reduction of work function to a value as small as 0.7 eV has been found for InSb.[80] Interestingly, if the FE current is re-calculated according to Equation (2) without introducing any surface roughness effect and assuming $\phi = 0.7$ eV on the effective emitting area, we obtain the real current intensity measured in the experiment and it corresponds to a field enhancement factor β≈3, which is the enhancement that results from the simulation (see Figure 6b).

## 4. Conclusion

We have performed extensive investigation of field emission properties of InSb NWs. The use of a metallic tip-anode, with curvature radius of 100 nm, collected the current emitted from a single NW. Field emission parameters, such as turn-on field and field enhancement factor showed non-monotonic dependence when the cathode-anode separation distance was is in the nanometer range (between 500 nm and 1300 nm). We demonstrated that an InSb NW is a very stable emitter, with field emitted current density as high as $10^4$ A/cm$^2$ and excellent time stability with fluctuations below 1% (of the order of $10^{-11}$ A), crucial parameters for device exploitation.

**Conflict of Interest**

The authors declare no conflict of interest.